\documentclass[aip,reprint]{revtex4-1}

\usepackage{lineno,hyperref}
\usepackage{color}
\usepackage{graphicx}% Include figure files
\usepackage{dcolumn}% Align table columns on decimal point
\usepackage{bm}% bold math
\usepackage[english]{babel}
\usepackage[utf8x]{inputenc}
\usepackage{siunitx}
\usepackage{amsmath}
\draft % marks overfull lines with a black rule on the right

\begin{document}

%\preprint{}

\title{Development of Yttrium alloy ion source and its application in nanofabrication} %Title of paper



\author{Nadezhda Kukharchyk}
\email[]{nadezhda.kukharchyk@physik.uni-saarland.de}
%\thanks{}
\altaffiliation{Universität des Saarlandes}
\affiliation{Experimentalphysik, Universität des Saarlandes, D-66123 Saarbrücken, Germany}
\affiliation{Angewandte Festkörperphysik, Ruhr-Universität Bochum, D-44780 Bochum, Germany}

\author{Ronna Neumann}
\affiliation{Angewandte Festkörperphysik, Ruhr-Universität Bochum, D-44780 Bochum, Germany}

\author{Swetlana Mazarov}
\affiliation{Angewandte Festkörperphysik, Ruhr-Universität Bochum, D-44780 Bochum, Germany}

\author{Pavel Bushev}
\affiliation{Experimentalphysik, Universität des Saarlandes, D-66123 Saarbrücken, Germany}

\author{Andreas D. Wieck}
\affiliation{Angewandte Festkörperphysik, Ruhr-Universität Bochum, D-44780 Bochum, Germany}

\author{Paul Mazarov}
\affiliation{Raith GmbH, Konrad-Adenauer-Allee 8, 44263 Dortmund, Germany}


\date{\today}

\begin{abstract}
	We present a new YAuSi Liquid Metal Alloy Ion Source (LMAIS), generating focused ion beams of yttrium ions, and its prospective applications for nanofabrication, sample preparation, lithographic and implantation processes. Working parameters of the AuSiY LMAIS are similar to other gold-silicon based LMAIS. We found anomalously high emission current of triple charged Yttrium ions. Influence of Yttrium implantation on optical qualities of the implanted ion-ensembles is shown in luminescence of co-implanted Erbium ions.
	
\end{abstract}


\maketitle %\maketitle must follow title, authors, abstract and \pacs


%PLAN OF THE PAPER

\section{Introduction and motivation}

Focused ion beams (FIB) have a lot of applications today, which vary from micro- and nano-machining of different structures \cite{nellen2006,sridhar2012,lacour2005,bischoff2016} to creation of robust and scalable quantum systems \cite{kukharchyk2014,probst2014,kukharchyk2016,siyushev2014,kornher2016,zhong2015}. 
In latest quantum information research, ion implantation became an irreplaceable tool in preparation of spin-ensembles in solid state matrix \cite{pezzagna2010,meijer2006,babinec2010}, which in particular relates to rare-earth (RE) spin-ensembles \cite{kukharchyk2014,siyushev2014,probst2014,xia2015}. Stable gold-silicon LMAIS like AuSiEr, AuSiCe or AuSiPr were repeatedly used in some of these experiments.

With an application of FIBs for spin ensembles preparation, concomitant ion-milling methods are more often required \cite{siyushev2014,zhong2015,babinec2010}. Typically, this micromachining or milling is done by Ga-ion beams. However, the usage of Ga-beam for milling requires after-etching of milled structures in order to prevent an influence from unavoidably implanted Ga ions \cite{zhong2015}.

Standard host-substrates for rare-earth ions are non-Ga-containing. These typically are Y$_2$SiO$_5$, YVO$_4$, YAlO$3$, Y$_3$Al$_5$O$_{12}$ etc. Ga-incorporation and thus the above mentioned etching step can be avoided if using Yttrium or Silicon ions for milling. The listed above substrates all have Yttrium ions in their crystalline matrix, and therefore milling of such materials with Yttrium ions could be more desirable and even profitable for RE ions properties. 

In this letter, we report on a development of AuSiY LMAIS and its prospective properties in micromachining and substrate nanoformation. First, we discuss the fabrication procedure and working parameters of the source. Then, we demonstrate its applications in preparation of ion- and spin-systems.

\section{Fabrication of the AuSiY LMIS}
The AuSiY LMIS was fabricated in the same procedure as described by Melnikov et al \cite{melnikov2002}. The emitter was fabricated of 0.2\,mm Tungsten wire in a common hairpin design. The heating wires were covered by ceramic glue in order to prevent creeping up and wetting of the heating wire with the alloy \cite{melnikov2002}. Fabrication of the needle was made according to the technique described by Prewett and Mair \cite{prewett1991}. 

A preliminary mixture of Au (78.4\%), Si (11.6\%) and Y (10\%) (atomic\%) was melted in a vacuum furnace. The LMAIS was filled by dipping it into this molten alloy under high vacuum conditions ($<10^{-5}$\,Pa). The alloy melting point is approximately 380$^\circ$C.

Working parameters of the source were first verified in a test-chamber (called  `emitter-maker'), and after that the source was tested in an EIKO 100 FIB system under ultra high vacuum (UHV) conditions ($\sim10^{-7}$\,Pa). Expected emission lifetime of the fabricated source is 1000 hours.

\section{Experimental results}
\subsection{Emission parameters}
\begin{figure}
	\includegraphics[width=\columnwidth]{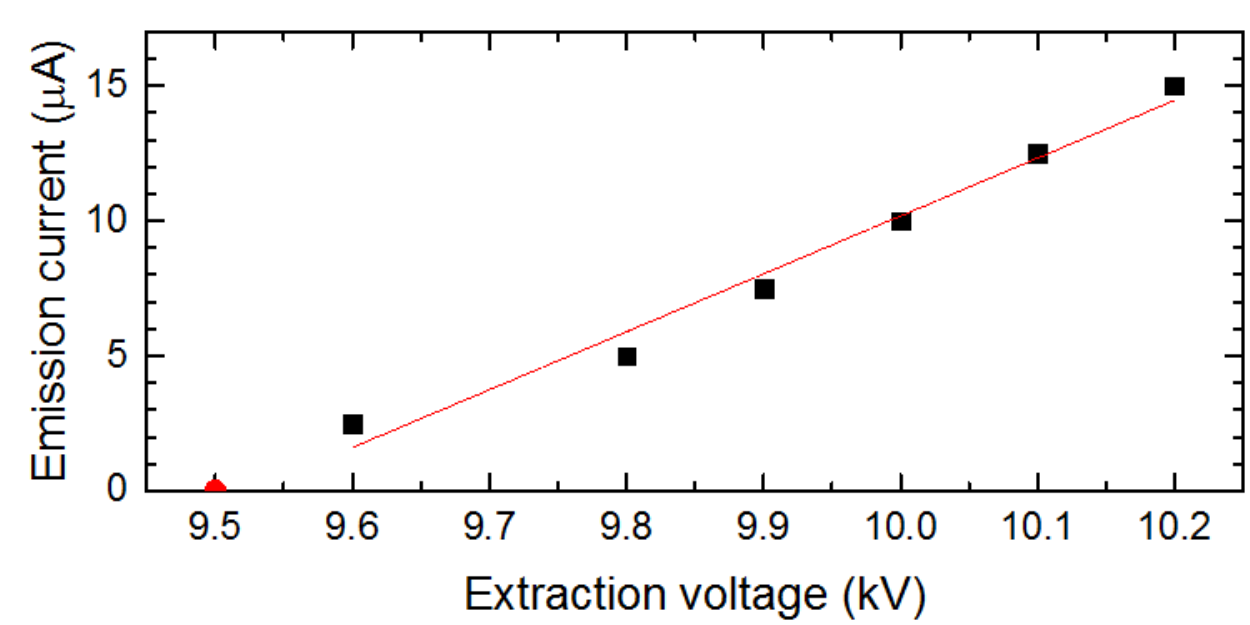}
	\caption{Current-voltage characteristic of the AuSiY source measured in the FIB machine. The I-V curve slope is equal to $\simeq 43\,\mu A/kV$. the heating current was 2.6\,\si{\ampere}. Below the threshold voltage, no emission current was observed.}.
	\label{fig1}
\end{figure}
Measured in EIKO 100 FIB, the current-voltage characteristic of AuSiY LMIS is shown in Fig.\,\ref{fig1}. The emission current was stable during operation times up to 8 hours. The slope of the I-V curve is equal to 43\,$\mu A/kV$, which is similar to other Au-based ternary alloys \cite{melnikov2002,bischoff2016}.

\subsection{Mass-spectrum}
\begin{figure}
	\includegraphics[width=\columnwidth]{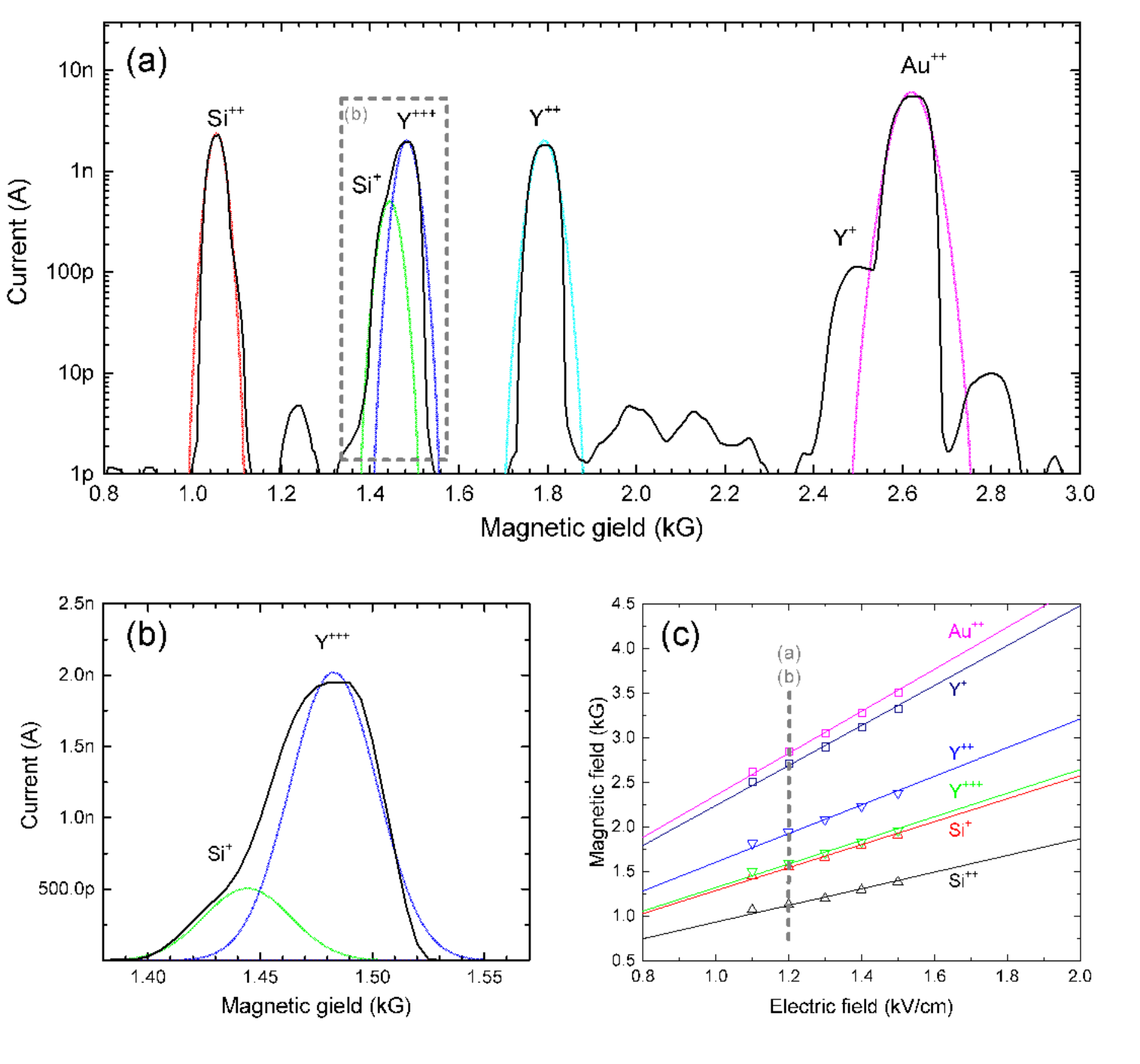}
	\caption{\textbf{(a)} Mass-spectrum of the AuSiY LMIS at the electric field of the E$\times$B mass-filter equal to 1.5\,kV/cm by accelerating voltage of 100\,kV and with a total source emission current of 15\,\si{\micro}A. \textbf{(b)} Overlapped peaks of Si$^+$ and Y$^{+++}$, magnified from (a). \textbf{(c)} Change of the ion peaks positions in the magnetic field with change of the electric field. The dotted lines correspond to the positions in the spectra (a) and (b).}
	\label{fig2}
	
\end{figure}

The mass-spectrum of the AuSiY LMIS was measured in the FIB with help of the Wien mass-filter at several electric field values. The resolution of the mass-filter was $m/\Delta m = 60$. Mass-spectrum for magnetic fields from 1\,kG to 4\,kG at 1.5\,kV/cm is shown in Fig.\,\ref{fig2}\,(a).
From the superposition of Electric force and Lorentz force, passage of the ions through the filter is given by 

\begin{equation}
B_{WF} = E_{WF}\,\sqrt{\frac{m\,u}{2E_{Acc}\,n}} \sim \sqrt{\frac{m}{n}},
\label{eq_WF}
\end{equation}
where $B_{WF}$ and $E_{WF}$ are respectively the magnetic and electric fields applied by the Wien filter; $m$ is the mass of the passing ion and $u$ is the atomic mass unit; $E_{Acc}$ is the acceleration energy of the ion; $n$ is the ion-charge. 
From \eqref{eq_WF}, we derive the order of ions in the mass-spectrum. 

One can see, that the Si$^{+}$ and Y$^{+++}$ peaks overlap as their mass-to-charge ratios take very close values: $B_{Si^{+}} \sim 28$ and $B_{Y^{+++}} \sim 29.7$, see Fig.\,\ref{fig2}\,(b). Thus, the first peak belongs to Si$^{+}$ and the second one to Y$^{+++}$. The current of Y$^{+++}$ ions is approximately three times higher as the current of Si$^{+}$, which results in a similar number of ions forming the peaks. 
In Fig.\,\ref{fig2}\,(c), positions of the mass-peaks are given in dependence on the $E \times B$ passage condition (\eqref{eq_WF}).

\begin{figure}
	\includegraphics[width=\columnwidth]{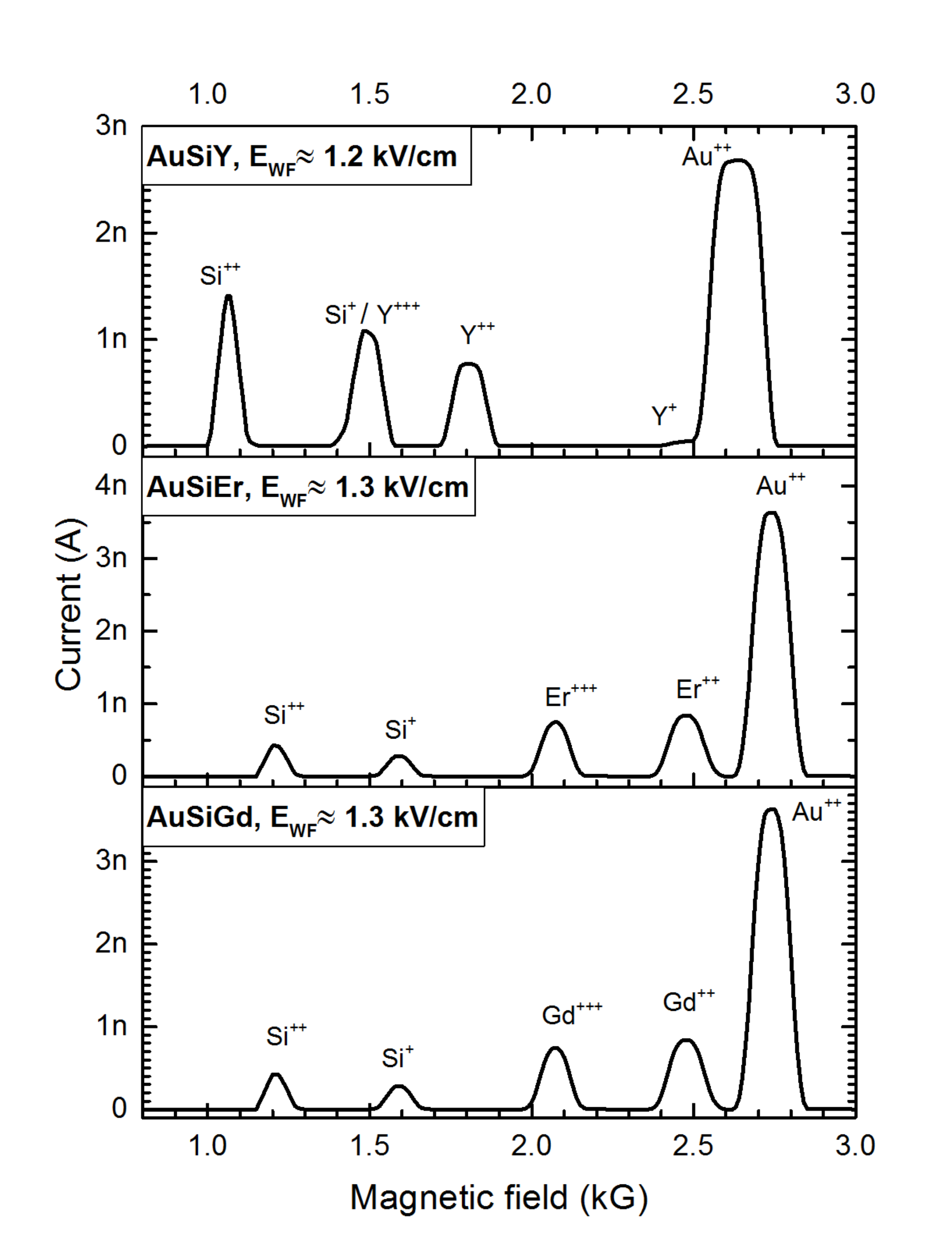}
	\caption{Comparison of the mass-spectra of several AuSiRE LMIS. The mass-spectra of AuSiEr and AuSiGd were taken at the electric field of 1.3\,kV/cm.}
	\label{fig3}
	
\end{figure}

The mass-spectrum of the AuSiY LMIS is similar to mass-spectra of other rare-earth-containing ion sources. To compare, we provide the mass-spectra of AuSiEr and AuSiGd, shown in Fig.\,\ref{fig3}. The mass-spectra of AuSiEr and AuSiGd were taken at the electric field of 1.3\,kV/cm, as noted in the legend. Relative positions of the peaks vary due to different ion masses, however relative intensities of the RE mass-peaks are similar for different sources.   

The sputter yield of SiO$_2$ by bombarding with Y$^{++}$ ions at 10\,kV was measured by normal (90$^{\circ}$) incidence of the ion beam onto the surface and it equals 60 atoms/ion.

\section{Applications}

\subsection{Surface milling}
The AuSiY ion-source can be used for milling various photonic structures replacing a Ga-ion-source for the substrates as Y$_2$SiO$_5$ (YSO), YAlO$_3$, YVO$_4$, Y$_3$Al$_5$O$_{12}$ etc. Y and Si ions can be well-separated in the mass-spectrum and Y$^{++}$ ion can be selected. For the materials like Y$_2$SiO$_5$, mixed peak Si$^+$/Y$^{++}$ can be used without separation of the Y and Si ions. This will allow to avoid contamination of the surface layers with gallium ions and the necessity to use etching methods afterwards. Thus, it leads to a more clean fabrication process.

\subsection{Co-doping effect}
\begin{figure}
	\includegraphics[width=\columnwidth]{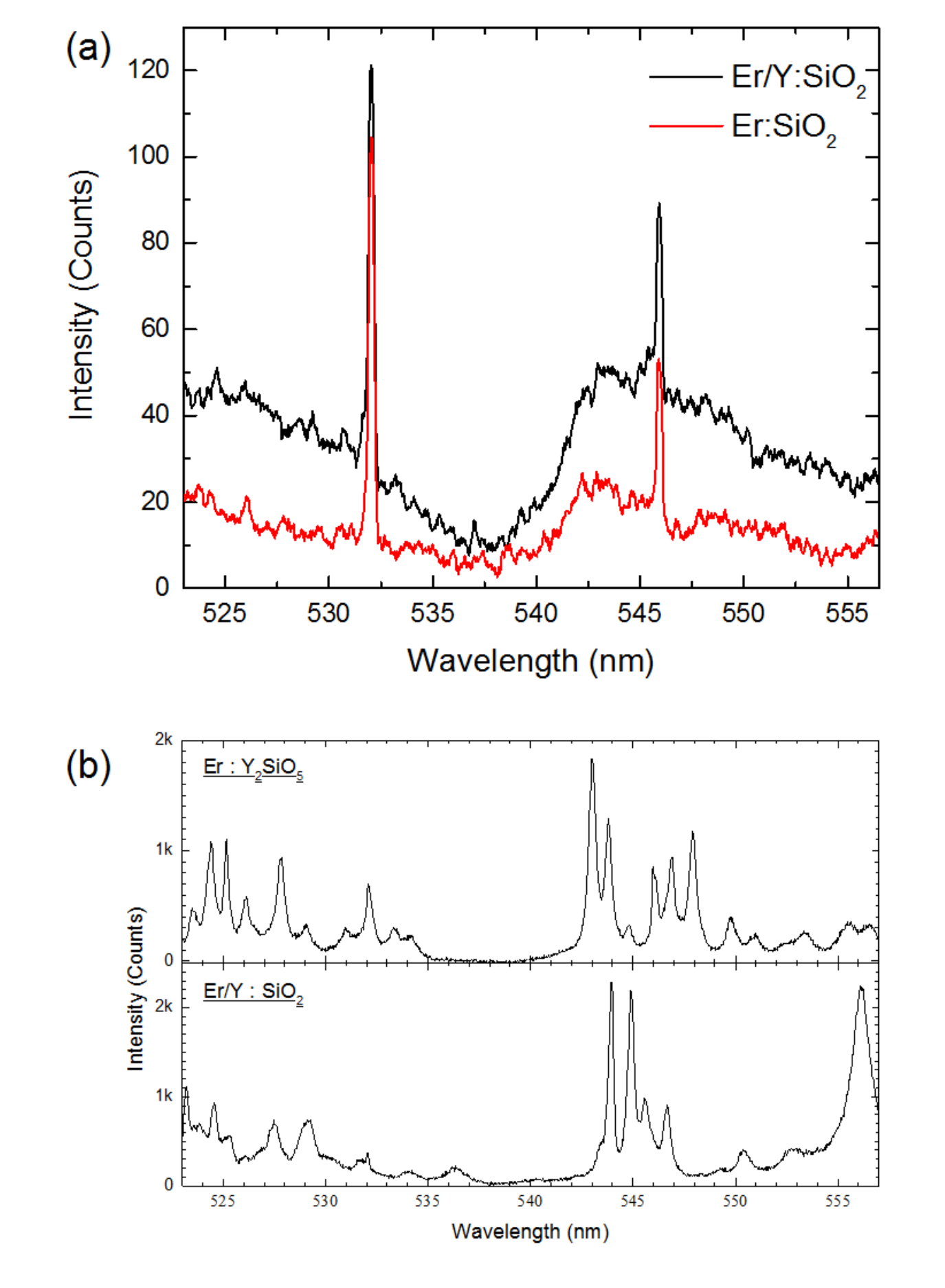}
	\caption{Co-doping with Yttrium of SiO$_2$ crystals. \textbf{(a)} Erbium luminescence in SiO$_2$ with and without co-dopes Yttrium ions. In the presence of Yttrium, the intensity of Erbium luminescence is twice increased. 
		\textbf{(b)} Structured luminescence of Er$_x$Y$_y$(SiO$_2$)$_z$ nanoclusters or nano-particles, compared to the Er:Y$_2$SiO$_5$ luminescence \cite{kukharchyk2016}.}
	\label{fig4}	
\end{figure}
We have implanted SiO$_2$ substrates with Erbium ions, part of which was co-implanted with Yttrium ions. A comparison of the following luminescence is shown in Fig.\,\ref{fig4}\,(a). The samples were excited with 638\,nm by an upconversion process in a confocal regime, and $^4$S$_{3/2}$/$^2$H$_{11/2}\rightarrow^4$I$_{15/2}$ transitions of Erbium were observed. The Yttrium fluence was equal to $10^{15}$\,cm$^{-2}$, and the Erbium fluence - to $4\times10^{13}$\,cm$^{-2}$. It can be seen from the spectra that the intensity of Erbium luminescence is twice enhanced in the presence of Yttrium ions.

\subsection{Nano-crystal formation}
In numerical works, chemical methods of forming YSO nanoparticles out of SiO$_2$ and Y$_2$O$_3$ \cite{wann2009,wang2010,marinovi2007,taghavinia2004} were demonstrated. We would like to emphasize here the creation of YSO nanoparticles in a porous SiO$_2$ matrix, as described by Taghavinia et al \cite{taghavinia2004}. There, a porous SiO$_2$ matrix is soaked with Yttrium and Europium solutions, dried out and annealed. In the result, Eu-doped YSO nanoparticles are formed in the SiO$_2$ matrix.

We propose, a similar result can be reached with implantation of Yttrium and Europium (Erbium, other rare earth) into the SiO$_2$ substrate. Or similarly, one can use Al$_2$O$_3$ substrates to form Y$_3$Al$_5$O$_{12}$ (YAG) nanocrystals. These two methods would require a proper Yttrium fluence and annealing procedure. 

For a demonstration, we have doped SiO$_2$ substrates with Erbium ($\sim 4\times10^{13}$\,cm$^{-2}$) and Yttrium ($\sim 10^{15}$\,cm$^{-2}$) and annealed those in air for several hours. Most of the sample area had broad luminescent peaks, as shown in Fig.\,\ref{fig4}\,(a), which come from variations in crystallographic symmetries of the SiO$_2$-glass (see for comparison luminescence of annealed/non-annealed Er:YSO in Kukharchyk et al. \cite{kukharchyk2014}). However, spots with defined repetitive symmetries were formed; this can be seen from the optical spectrum in Fig.\,\ref{fig4}\,(b). To compare, a spectrum of Erbium implanted Y$_2$SiO$_5$ is shown in Fig.\,\ref{fig4}\,(b) as well. It cannot be concluded here if a different phase of Y$_2$SiO$_5$ or even Y$_2$SiO$_7$ was formed. Nevertheless, it demonstrates that with such a local implantation, nano-particles/-clusters of rare-earth-doped YSO can be created and applied in various modern research.

\section{Conclusion}

We have reported on a fabrication of a stable AuSiY LMAIS. AuSiY LMAIS has a high Yttrium-ion current and high sputter rate. Emitting properties of the source are similar to the other AuSiRE ion-sources. In the mass-spectrum, overlapping Si$^+$/Y$^{+++}$ peaks are well separated already at 1.5\,kV/cm.
AuSiY LMAIS can be applied in both micromachining and fabrication of luminescent nanoparticles for quantum optics and quantum information research.

\section{Acknowledgement}
Authors would like to thank Roman Kolesov, Kangwei Xia and Jörg Wrachtrup for the possibility of optical measurements.

% If in two-column mode, this environment will change to single-column format so that long equations can be displayed. 
%\begin{widetext}
%$$\mbox{put long equation here}$$
%\end{widetext}

% \begin{figure}
% \end{figure}

% \begin{table}
% \end{tabular}
% \end{table}

%\begin{acknowledgments}
% Put your acknowledgments here.
%\end{acknowledgments}

\bibliographystyle{apsrev4-1}
\bibliography{library}

\end{document}